\documentclass[11pt,showpacs]{revtex4}
\begin{document}

\preprint{}

\title{On Quantum Radiation in Curved Spacetime }

\author{She-Sheng Xue}

\email{xue@icra.it}

\affiliation{ICRA, INFN and
Physics Department, University of Rome ``La Sapienza", 00185 Rome, Italy}



\begin{abstract}
In the context of quantum field theories in curved 
spacetime, we compute the effective action of the 
transition amplitude from vacuum to vacuum in the presence
of an external gravitational field. The imaginary part of resulted effective action determines 
the probability of vacuum decay via quantum tunneling process, giving the rate and spectrum
of particle creations. We show that gravitational field polarizes vacuum and discretizes its spectrum
for such a polarization gains gravitational energy. On the basis of gravitational vacuum polarization, 
we discuss the quantum origin of vacuum decay in curved spacetime as pair-creations of particles and 
anti-particles. The thermal spectrum of particle creations is attributed to (i) the CPT invariance of 
pair-creations(annihilations) from(into) vacuum and (ii) vacuum acts as a reserve with the temperature 
determined by gravitational energy-gain. 
\end{abstract}


\maketitle

\section{\it Introduction.}

The issue of quantum field theories of elementary particles in curved spacetime has shown a 
tremendously important role of
understanding quantum phenomenon of particle creations in the presence of gravitational field. 
The Hawking radiation\cite{hawking75} is one of such phenomena occurring
around black hole's horizon. It has advocated a numerous studies in the last three 
decades\cite{book}. Briefly and broadly speaking, in these studies there are two
approaches to the Hawking radiation. In the first, one considers an incoming and outgoing wave 
destoried in a collapse geometry, and
the Hawking radiation can explicitly and implicitly be approached by an appropriate boundary
condition. In the second, one treats a black hole immersed in a thermal bath, implying the
Hawking radiation from black holes by the detail balance. These approaches do not very 
directly reveal the genuine quantum process taking place for the source of 
the Hawking radiation, 
although, the Hawking radiation is heuristically considered as pair-creations of particles and 
antiparticles by a quantum
effect of tunneling\cite{pair}, analogous to the Euler-Heisenberg-Schwinger 
process in QED\cite{sw}. In recent studies\cite{thooft,wilczk00},
the semi-classical WKB-method for quantum tunneling process is adopted to reveal 
the quantum tunneling nature of the Hawking radiation and its back 
reaction\cite{wilczk00}.
For explicitly revealing the genuine origin of quantum radiation in curved spacetime, it is
essential to analyze the effect in the context of quantum field theories. This is also very 
important for understanding microscopic origin of 
entropy in black hole thermodynamics and the problem of unitality.

In a path-integral framework of quantum field theories, we compute the effective action 
of transition amplitude from vacuum to vacuum in curved spacetime. 
With the Schwarzschild geometry, we obtain the imaginary part of effective action, which 
gives rise to the probability of vacuum decay via quantum tunneling process 
for pair-creations of 
particles and anti-particles. We discuss that 
this quantum emission is dynamically attributed to (i) quantum-field
fluctuations of positive- and negative-energy virtual particles in vacuum are 
polarized by external gravitational field, (ii) such a vacuum polarization gains gravitational energy
and (iii) effective mass-gap separating positive-energy particles from negative-energy particles is very small. 
The energy spectrum of vacuum is discretized by external gravitational field.
The spectrum of particle creations is 
determined by quantum emissions(absorptions) from(into) vacuum, in unit of quanta of the discrete 
energy spectrum of vacuum.   

\section{\it General formulation.}

We assume that the structure of spacetime is described by the pseudo-Riemannian metric $g_{\mu\nu}$
associated with the line element
\begin{equation}
ds^2=g_{\mu\nu}dx^\mu dx^\nu,\hskip0.5cm \mu,\nu=0,1,2,3,
\label{line}
\end{equation}
and the spacetime point is coordinated by $x=(x^0,x^i)=(t,\vec x)$.
The special geometrical symmetries of the spacetime $\cal S$ are described by using 
Killing vectors $\xi^\mu$, which are solutions of Killing's equation
\begin{equation}
{\cal L}_\xi g_{\mu\nu}(x)=0,\hskip0.5cm \xi_{\mu;\nu}+\xi_{\nu;\mu}=0,
\label{killing}
\end{equation}
where ${\cal L}_\xi$ is the Lie derivative along the vector field $\xi^\mu$, orthogonal to the 
spacelike hypersurface $\Sigma_t$ ($t=$constant) of the spacetime ${\cal S}$. 
A static observer ${\cal O}$ is at rest in this hypersurface $\Sigma_t$.
We consider quantum-field fluctuations interacting with curved spacetime.

In order to clearly illustrate physics content,
we first consider a complex scalar field $\phi$ in curved spacetime. 
The simplest coordinate-invariant action is given by ($\hbar=c=G=k=1$)
\begin{equation}
S = {1\over2}\int
d^4x\sqrt{-g}\Big[g^{\mu\nu}\phi_{,\mu}\phi^*_{,\nu}+(m^2+\xi {\cal R})
\phi\phi^*\Big],
\label{action}
\end{equation}
where $m$ is particle mass and ${\cal R}$ the Riemann scalar. 
The quantum scalar field $\phi$ can be in principle expressed in terms of
a complete and orthogonal basis of quantum-field states $u_k(x)$:
\begin{equation}
\phi(x)=\sum_k\Big(a_k u_k(x)+
a^\dagger_ku^*_k(x)\Big),\hskip0.1cm \left [a_k,a^\dagger_{k'}\right ]
=\delta_{k,k'}
\label{de}
\end{equation}
where $a^\dagger_k$ and $a_k$ are creation and annihilation operators
of the $k$-th quantum-field state $u_k(x)$. This quantum field state obeys the 
following equation of motion,
\begin{equation}
(\Delta_x + m^2+\xi {\cal R})u_k(x)=0,
\label{eq}
\end{equation}
and appropriate boundary conditions for selected values of $k$. In Eq.(\ref{eq}),
$\Delta_x$ is the Laplacian operator in curved spacetime:
\begin{equation}
\Delta_x = {1\over\sqrt{-g}}\partial_\mu\big[\sqrt{-g}g^{\mu\nu}\partial_\nu\big].
\label{laplace}
\end{equation}
In Eq.(\ref{de}), we assume that $u_k(x)$ are positive energy ($\omega$) states, 
satisfying
\begin{equation}
{\cal L}_\xi u_k(x)=-i\omega u_k(x), \hskip0.5cm \omega >0,
\label{killingv}
\end{equation} 
with respect to the timelike Killing vector field $\xi^\mu$ (\ref{killing}) associated to 
the static observer ${\cal O}$. 

We assume that in the remote past $(t\rightarrow -\infty)$, massive dust is uniformly
distributed that the spacetime is approximately flat. 
Quantum-field states $u_k(x)$ (\ref{eq}) in the remote past $(t\rightarrow -\infty)$
are asymptotically free states $\bar u_k(x)$, obeying Eq.(\ref{eq}) for  
$g_{\mu\nu}\simeq (1,-1,-1,-1)$. Then, the quantum scalar field $\phi(x)$ is an 
asymptotically free field in the hypersurface $\Sigma_{-\infty}$ of the spacetime ${\cal S}$:
\begin{equation}
\phi_{\rm in}(x)=\sum_k\Big(\bar a_k \bar u_k(x)+
\bar a^\dagger_k\bar u^*_k(x)\Big),\hskip0.1cm \left [\bar a_k,\bar a^\dagger_{k'}\right ]
=\delta_{k,k'}
\label{de1}
\end{equation}
where $\vec x\in \Sigma_{-\infty}$, $\bar a^\dagger_k$ and $\bar a_k$ are creation and annihilation operators
of the $k$-th asymptotically free quantum-field state $\bar u_k(x)$. 
Corresponding Lie derivative along the Killing vector (\ref{killing}) is $\partial_t$, 
positive energy states are $\bar u_k(x)$, satisfying Eq.(\ref{killingv}). 
Then we may construct the standard Minkowski space quantum vacuum state 
$|\bar 0,{\rm in}\rangle$:
\begin{equation}
\bar a_k |\bar 0,{\rm in} \rangle=0,\hskip0.5cm \langle \bar 0,{\rm in}|\bar a_k^\dagger=0.
\label{vd1}
\end{equation}
$|\bar 0,{\rm in} \rangle$ is an initial quantum vacuum state in the remote past 
$(t\rightarrow -\infty)$ with respect to the static observer ${\cal O}$.

In the remote future $(t\rightarrow +\infty)$, a massive object is formed and stable
that the spacetime is curved and stationary, described by the geometry $g_{\mu\nu}$. We assume that 
quantum-field states $u_k(x)$ (\ref{eq}) in the remote future $(t\rightarrow +\infty)$
are asymptotical states $\tilde u_k(x)$, obeying Eq.(\ref{eq}) for a stationary geometry 
$g_{\mu\nu}$. In the hypersurface $\Sigma_{+\infty}$ of the spacetime ${\cal S}$,
the asymptotical quantum scalar field $\phi$ in the remote future $(t\rightarrow +\infty)$ 
is expressed in terms of $\tilde u_k(x)$:
\begin{equation}
\phi_{\rm out}(x)=\sum_k\Big(\tilde a_k \tilde u_k(x)+
\tilde a^\dagger_k \tilde u^*_k(x)\Big),\hskip0.1cm \left 
[\tilde a_k,\tilde a^\dagger_{k'}\right ]=\delta_{k,k'}
\label{de2}
\end{equation}
where $\vec x\in \Sigma_{+\infty}$, $\tilde a^\dagger_k$ and $\tilde a_k$ are creation and annihilation operators
of the $k$-th quantum-field state $\tilde u_k(x)$. Corresponding Lie derivative along the Killing vector is $\xi^\mu$ (\ref{killing}), positive energy 
states are $\tilde u_k(x)$ satisfying Eq.(\ref{killingv}). Then we may construct the
quantum vacuum state $|\tilde 0,{\rm out}\rangle$:
\begin{equation}
\tilde a_k |\tilde 0,{\rm out}\rangle=0,\hskip0.5cm 
\langle \tilde 0,{\rm out}|\tilde a_k^\dagger=0,
\label{vd}
\end{equation}
in curved spacetime.
$|\tilde 0,{\rm out} \rangle$ is an final quantum vacuum state in the remote future 
$(t\rightarrow +\infty)$ with respect to the same static observer ${\cal O}$.
 
It is worthwhile to note that $\phi_{\rm out}(x)$($\{\tilde u_k(x)\}$) are not asymptotically free field(states),
instead they are asymptotical field(states) in the presence of external stationary gravitational field,
so that the final vacuum state $|\tilde 0,{\rm out}\rangle$ (\ref{vd}) is different from 
the initial vacuum state $|\bar 0,{\rm in}\rangle$ (\ref{vd1}). As will be seen, such a difference
is not only a unitary phase. The final vacuum state $|\tilde 0,{\rm out}\rangle$ (\ref{vd}), may not necessarily be measured as devoid of particles, in contrast to the initial vacuum state $|\bar 0,{\rm in}\rangle$ 
defined by Eq.(\ref{vd1}) relating to the asymptotically free field $\phi_{\rm in}(x)$.
In fact, as will be shown, the final vacuum state $|\tilde 0,{\rm out}\rangle$ is a quantum-field 
state of particle and antiparticle creations upon the initial vacuum state $|\bar 0,{\rm in}\rangle$. 
This indicates gravitational 
field interacting with quantum-field fluctuations of positive and negative energy states 
of the initial vacuum $|\bar 0,{\rm in} \rangle$, and the quantum scalar field evolves throughout 
intermediate states $\phi(\vec x,t)$ (\ref{de}) for $-\infty <t<+\infty$. We speculate that this 
evolution is adiabatic for the reasons that gravitational-field energy involved 
is very small and the evolution takes infinite time.  

To deal with all possible 
intermediate states, represented by $\phi(x)$ or $u_k(x)$ in Eq.(\ref{de}), for $-\infty <t<+\infty$, we use path-integral representation to study the transition
amplitude between the initial vacuum state and final vacuum state: 
\begin{equation}
\langle \tilde 0,{\rm out}|\bar 0,{\rm in}\rangle=
\int [{\cal D}\phi{\cal D}\phi^*]\exp(iS),
\label{action2}
\end{equation}
where 
\begin{equation}
\int [{\cal D}\phi{\cal D}\phi^*]=\Pi_{-\infty <t<+\infty}\Pi_{\vec x\in\Sigma_t}\int [d\phi(\vec x,t)\phi^*(\vec x,t)].
\label{measure}
\end{equation}
The intermediate states contributions to the transition amplitude (\ref{action2}) 
can be formally path-integrated,
\begin{equation}
\langle \tilde 0,{\rm out}|\bar 0,{\rm in}\rangle 
= {\det}^{-1}\left({\cal M}\right),\hskip0.2cm
{\cal M}\! =\!\Delta_x + m^2+\xi {\cal R}.
\label{m}
\end{equation}
This result is in terms of $\phi_{\rm in}$ (\ref{de1}) and $\phi_{\rm out}$ (\ref{de2}), which are not explicitly
written.
The effective action $S_{\rm eff}$ is defined as
\begin{equation}
S_{\rm eff}=-i\ln \langle \tilde 0,{\rm out}|\bar 0,{\rm in}\rangle,
\label{action1}
\end{equation}
which relates to the phase of the $S$-matrix transition from the initial vacuum state 
$|\bar 0,{\rm in}\rangle$ to the final vacuum state $|\tilde 0,{\rm out}\rangle$. We focus on 
the calculations of the effective action (\ref{action1}) in this article. 

In order to evaluate the path-integral (\ref{action2}) over all intermediate states $\phi(x)$ 
(\ref{de}), it is convenient to introduce operators $\hat X_\mu$ and $\hat K_\mu$ 
defined on the states $|x\rangle$ and $|k\rangle$:
\begin{equation}
\hat X_\mu|x\rangle=x_\mu|x\rangle;\hskip0.3cm \hat K_\mu|k\rangle=k_\mu|k\rangle.
\label{xk}
\end{equation}
They enjoy the canonical communication,
\begin{equation}
[\hat X_\mu,\hat K_\nu]=-ig_{\mu\nu}.
\label{cxk}
\end{equation}
The states $|x\rangle$ and $|k\rangle$ satisfy:
\begin{eqnarray}
\langle x| x'\rangle &=&\delta(x-x'),\hskip0.5cm 
\int dx|x\rangle \langle x|=1\nonumber\\
\langle k| k'\rangle &=& 2\pi\delta(k-k'),\hskip0.3cm  
\int dk|k\rangle \langle k|=1,
\label{basis}
\end{eqnarray}
and intermediate quantum-field state $u_k(x)$ can be represented as
\begin{equation}
u_k(x)=\langle x|k\rangle.
\label{uxk}
\end{equation}
Using these matrix notations, we write the operator ${\cal M}(\hat X,\hat K)$ (\ref{m}) 
as a hermitian matrix 
\begin{eqnarray}
{\cal M}_{k,k'}\! &=& \!\int dxdx'\langle k|x\rangle\langle x|{\cal M}(\hat X,\hat K)| x'
\rangle\langle x'| k' \rangle\nonumber\\
\! &=&\! \int dx u^*_k(x){\cal M}(x,\hat K)u_{k'}(x).
\label{mxk}
\end{eqnarray}
In this representation, diagonizing this hermitian matrix ${{\cal M}_{k,k'}}$, 
we formally compute the effective action $S_{\rm eff}$ given in Eqs.(\ref{action1},\ref{m}):
\begin{equation}
iS_{\rm eff}=- {\rm tr}\ln({\cal M})=-\int {d^4k\over (2\pi)^4}
\ln\lambda^2_k,
\label{zr}
\end{equation}
where the $\{\lambda^2_k\}$ is the diagonal element of the matrix (\ref{mxk})
\begin{equation}
\lambda^2_k=\int \sqrt{-g}d^4x \tilde u^*_k(x){\cal M}(x,\hat K)\tilde u_{k}(x),
\label{lambdat}
\end{equation}
in the phase space ($k$) of the quantum field states $\{\tilde u_{k}(x)\}$.
The operator $\lambda^2_k$ (\ref{lambdat}) and the number of quantum-field states 
$\int \sqrt{-g}d^4xd^4k/(2\pi)^4$ are invariant in
arbitrary coordinate systems, later 
is the Liouville theorem for the phase-space invariance.
In Eqs.(\ref{zr}) and (\ref{lambdat}), the term
that is independent of non-trivial geometry $g_{\mu\nu}$ has been dropped.

\section{\it Spectrum of vacuum.}

At the remote future $(t\rightarrow +\infty)$, we assume the geometry of the spacetime 
``'${\cal S}$' outside of a massive object $M$
($r>2M$) is stationary and spherical, e.g., the Schwarzschild geometry,
\begin{eqnarray}
ds^2\!&=&\!g_{tt}dt^2\!+\!g_{rr}dr^2\!+\!r^2d\Omega,\nonumber\\
-g_{tt}\!&=&(g_{rr})^{-1}=g(r)\!\equiv\! (1\!-\!{2M\over r})
\label{sg}
\end{eqnarray}
where $\Omega$ is the spherical solid angle and $r,\theta,\phi,t$
are the Schwarzschild coordinates.
The static observer located at ${\cal O}$ at $x = (t,r,\Omega)$, whose 
four velocity $u_\mu$ and Killing vector $\xi_\mu$ are given by, 
\begin{equation}
u_\mu = (g^{1\over2}_{tt}(r),0,0,0),\hskip0.5cm
\xi_\mu  = (g_{tt}(r),0,0,0),
\label{o}
\end{equation}
which are orthogonal to the spacelike hypersurface $\Sigma_t$ ($t=$constant, 
$\vec x=(r.\Omega)\in\Sigma_t $). We will respectively
discuss the static observers locates at finite radius ($r=r^*<\infty$) and at infinity 
($r=r^*\rightarrow\infty$) of the hypersurface $\Sigma_t$. 
The Riemann scalar ${\cal R}=0$. The Laplacian operator 
(\ref{laplace}) is given by:
\begin{eqnarray}
\Delta_x &=& g^{tt}{\partial^2\over \partial t^2}+{1\over r^2}{\partial\over \partial r} 
r^2g^{rr}{\partial\over \partial r}
+{\hat L^2\over r^2}\nonumber\\
&=&-g^{-1}(r){\partial^2\over \partial t^2} +g(r)\left({\partial^2\over \partial r^2}
+{2\over r}{\partial\over \partial r}\right)\nonumber\\
&+&{2M\over r^2}{\partial\over \partial r}
+{\hat L^2\over r^2},
\label{laplace1}
\end{eqnarray}
where $\hat L^2$ is the angular momentum operator.

The final vacuum state $|\tilde 0,{\rm out} \rangle$ of this geometry is usually denoted 
as the Schwarzschild vacuum $|0_S\rangle$. The appropriate basis of asymptotical quantum field 
$\phi_{\rm out}(x)$ (\ref{de2}) is chosen as 
\begin{equation}
\tilde u_k(x)= \langle t,r,\theta,\phi|\omega,k_r,l,m\rangle =R_{l\omega}(r)Y_{lm}
(\theta,\phi)e^{i\omega t},
\label{sphere}
\end{equation} 
where $k$ indicates a set of quantum numbers $(\omega,k_r,l,m)$. $\omega$ 
is the energy-spectrum and 
$Y_{lm}(\theta, \phi)$ is the standard spherical harmonic function: 
$\hat L^2Y_{lm}(\theta, \phi)=l(l+1)Y_{lm}(\theta, \phi)$. The radial function 
$R_{l\omega}(r)$ obeys the following differential equation for $r>2M$,
\begin{equation}
\left[\omega^2-g^2(r)\hat k_r^2+i{2M\over r^2}g(r)\hat k_r-g(r)V_l(r)\right]R_{l\omega}(r)=0,
\label{eq2}
\end{equation}
where the hermitian radial momentum operator,
\begin{equation}
\hat k_r={1\over ir}{\partial\over\partial r}r,\hskip0.3cm (\hat k_r)^2
=-\left({\partial^2\over\partial r^2}+{2\over r}{\partial\over\partial r}\right),
\label{pr}
\end{equation}
and potential
\begin{equation}
V_l(r)={l(l+1)\over r^2}+{2M\over r^3}+m^2.
\label{p}
\end{equation}
Eq.(\ref{eq2}) is exactly equivalent to the Regge and Wheeler equation. The radial 
function $R_{l\omega}(r)$ is orthogonal
and asymptotically behaves as the Hankel function $h_{l\omega}(r)$ for $r\gg 2M$.

The matrix operator ${\cal M}$ in Eq.(\ref{lambdat}) is given by,
\begin{equation}
{\cal M}(r,\hat k_r)=\omega^2-g^2(r)\hat k_r^2+i{2M\over r^2}g(r)\hat k_r-g(r)V_l(r).
\label{matrix}
\end{equation}
Eqs.(\ref{sphere}-\ref{matrix}) define a complex eigen-value problem to find the energy 
spectrum of the vacuum state (\ref{vd}) in an external gravitation field. 
The imaginary part of the  
operator ${\cal M}(r,\hat k_r)$ (\ref{matrix}), giving rise to an imaginary part of the 
effective action (\ref{action1}), implies that the gravitating quantum-field system 
(\ref{action}) be energetically unstable and quantum-field tunneling could occur, 
leading to the productions of particles and antiparticles.
 
Up to an irrelevant constant in Eq.(\ref{zr}), the diagonal matrix $\lambda^2_k$ (\ref{lambdat}) is given by:
\begin{eqnarray}
\lambda^2_k&\equiv &N^{-1}_k\int \sqrt{-g}d^4x \tilde u_k(x)(\Delta_x + m^2) \tilde u^{*}_{k}(x),\nonumber\\
&=&N^{-1}_k\int dr r^2R^*_{l\omega}(r){\cal M}(r,\hat K_r)R_{l\omega}(r),
\label{mxk10}
\end{eqnarray}
where the normalization factor is
\begin{equation}
N_k=\int dr r^2R^*_{l\omega}(r)R_{l\omega}(r).
\label{norma0}
\end{equation}
The elements of the diagonal matrix $\lambda^2_k$ (\ref{mxk10}) is related to the 
inverse propagator (spectrum)
of particles in the external gravitation field. We write Eq.(\ref{mxk10}) as,
\begin{equation}
\lambda^2_k=\omega^2-\kappa_k^2- m^2_k+i2M\epsilon_k,
\label{ipro0}
\end{equation}
where we define the average values $\kappa_k^2$, $\epsilon_k$ and effective mass-gap 
$m_k$ of the quantum field state $R_{l\omega}(r)$:
\begin{eqnarray}
\kappa_k^2 &\equiv& N^{-1}_k \int dr r^2R^*_{l\omega}(r)g^2(r)\hat k^2_r R_{l\omega}(r);
\label{kr20}\\
m^2_k&\equiv& N^{-1}_k\int dr r^2R^*_{l\omega}(r)g(r)V_l(r)R_{l\omega}(r);
\label{mb0}\\
\epsilon_k &\equiv& N^{-1}_k\int dr r^2R^*_{l\omega}(r)
{g(r)\over r^2}\hat k_r R_{l\omega}(r).
\label{epsilon0}
\end{eqnarray}

By the definition (\ref{epsilon0}), the sign of $\epsilon$ is determined by the 
sign of the radial momentum $k_r$ of quantum field state $R_{l\omega}(r)$,
\begin{equation}
k_r \equiv N^{-1}_k\int dr r^2R^*_{l\omega}(r)\hat k_r 
R_{l\omega}(r).
\label{kr0}
\end{equation}
$\epsilon_k >0$ indicates that the radial momentum of the state $R_{l\omega}(r)$ 
is positive ($k_r>0$), standing for outgoing states of positive-energy particles 
$\omega>0$. Whereas, 
$\epsilon_k <0$ indicates that the radial momentum of the state $R^*_{l\omega}(r)$ 
is negative ($k_r<0$), standing for incoming states of negative energy particles 
$\omega<0$. This is in accordance with Feynman's  
$\omega\rightarrow\omega\pm i2M\epsilon_k$ prescription for 
particles(+) and antiparticles(-), latter are outgoing states with positive energy
$(\omega>0)$ and positive radial momentum, traveling backward in time. In this 
prescription, we have positive energy $\omega>0$ and positive radial momentum 
$|k_r|$ (\ref{kr0}) for both particles and antiparticles.

\section{\it Effective action.}

Given $\lambda^2_k$ (\ref{mxk10}), we can write the effective action (\ref{zr}) as,
\begin{equation}
iS_{\rm eff}=-\int {d^4k\over (2\pi)^4}
\ln(\lambda^2_k)=- \int {d\omega dk_r\over (2\pi)^2} \sum_{l,m}
\ln(\lambda^2_k).
\label{zr00}
\end{equation}
Using the identity:
\begin{equation}
\ln{a\over b}=\int_0^\infty {ds\over s}\left(e^{is(b+i\epsilon)}-e^{is(a+i\epsilon)}\right).
\label{id}
\end{equation}
we are able to write the effective action (\ref{zr00})
\begin{equation}
S_{\rm eff}=-i\int {d\omega dk_r\over (2\pi)^2} \sum_{l,m}
\int_0^\infty {ds\over s}e^{is(\lambda^2_k+i\epsilon)}-(M=0),
\label{zr10}
\end{equation}
where the expression of the second part indicated by $(M=0)$ is the same as the expression of
the first part with $M=0$. The logarithmatic function in Eq.(\ref{zr00}) is represented by 
an $s$-integration in Eq.(\ref{zr10}) and infrared convergence at $s\rightarrow 0$ is 
insured by $i\epsilon$ prescription ($\epsilon\rightarrow 0$). 

We begin with the computation of summing over quantum numbers ``$l,m$'' of angular momenta 
in Eq.(\ref{zr10}). 
Eqs.(\ref{p},\ref{matrix},\ref{mxk10}) show that $\lambda^2_k$ (\ref{mxk10}) 
depends on ``$l$'' in terms of $l(l+1)/r^2$ and  $R_{l\omega}(r)$. We find that 
the $l$-dependence is dominated by the quadratic term $l(l+1)/r^2$, 
in contrast with this, $R_{l\omega}(r)$ is a smooth function 
in varying ``$l$''. Thus, to achieve the leading contribution to the effective action (\ref{zr10}), 
we approximate the quantum-field state $R_{l\omega}(r)$ in $\lambda^2_k$ (\ref{mxk10}) as:
\begin{equation}
R_{l\omega}(r)\simeq R_{l^*\omega}(r)
\label{app0}
\end{equation}
where $l^*$ is a particular value of angular quantum number, for an example, 
$l^*=0$ for the spherically 
symmetric state. Then, in Eq.(\ref{zr10}) we approximately have,
\begin{equation}
\sum_{l,m}e^{is(\lambda^2_k+i\epsilon)}\simeq
e^{is(\Lambda^2_k+i\epsilon)}\sum_{l,m}e^{-is\beta_\omega l(l+1)},
\label{suml0}
\end{equation}
where 
\begin{eqnarray}
\Lambda^2_k &=& \lambda^2_k + \beta_k l(l+1)\label{lambda0}\\
\beta_k &=& N_k^{-1}\int dr r^2|R_{l^*\omega}(r)|^2\left({g(r)\over r^2}\right).
\label{app10}
\end{eqnarray}
In addition, we introduce continuous and dimensionless transverse momenta $\vec k_\perp$ 
and $k^2_\perp=l(l+1)$ so that,
\begin{equation}
\sum_{lm}e^{-is\beta_k l(l+1)}\simeq 4\pi \int{d^2k_\perp\over (2\pi)^2}
e^{-is\beta_k k^2_\perp}={1\over is\beta_k},
\label{angu0}
\end{equation}
where $\beta_k$ is analytically continued to complex value and ${\rm Im}(\beta_k)<0$. 
As a result, we obtain the effective action (\ref{zr10}),
\begin{equation}
S_{\rm eff}\simeq -\int {d\omega dk_r\over (2\pi)^2} {1\over\beta_k}\int_0^\infty
{ds\over s^2}e^{is( \Lambda^2_k+i\epsilon)}-(M=0).
\label{seff0}
\end{equation}

In order to compute the integration over ``$s$'' in Eq.(\ref{seff0}), we introduce a complex
variable $z=-1+\delta$ ($|\delta| \rightarrow 0 $), and use the following integral
representation of the $\Gamma(z)$-function by an analytical continuation 
for ${\rm Im}(\alpha) >0$:
\begin{equation}
\int_0^\infty e^{i\alpha s}s^{z-1}ds=(-i\alpha)^{-z}\Gamma(z).
\label{int}
\end{equation}
This analytical continuation is equivalent to the analytical continuation of 
dimensionality of the transverse momentum-space,
\begin{equation}
\int{d^2k_\perp\over (2\pi)^2}\rightarrow \int{d^{2+\delta}
k_\perp\over (2\pi)^{2+\delta}},
\label{intana}
\end{equation}
in Eq.(\ref{angu0}).
In the neighborhood of singularity, where $|\delta| \rightarrow 0$ and 
$z\rightarrow -1$ in Eq.(\ref{int}), we have 
\begin{eqnarray}
\Gamma(z)&=&-{1\over\delta}-1+\gamma+O(\delta),\nonumber\\
\alpha^{-z}&=&
\alpha e^{-\delta\ln{\alpha\over\mu^2}},\nonumber\\
(-i)^{-z}&=&-ie^{i\delta(-{\pi\over2}+2\pi n')},\nonumber\\  
n'&=&0,1,2,3\cdot\cdot\cdot
\label{ana}
\end{eqnarray}
where the Euler constant $\gamma\simeq 0.577$ and $\mu$ is an arbitrary scale of 
dimensional transmutation in the dimensional regularization. 
The singular term $1/\delta$ corresponds to the ultra-violet divergence of 
summing over $(l,m)$ in Eq.(\ref{zr10}).

Analogously to the treatment of dimensional and $\zeta$-functional
regularization schemes, using Eqs.(\ref{int},\ref{ana}) to calculate integration
over ``$s$'' in Eq.(\ref{seff0}) and keeping up to the
order $O(\delta^\circ)$, we cast the effective action Eq.(\ref{seff0}) to be:
\begin{eqnarray}
\!S_{\rm eff}\!&=&\!-\int {d\omega dk_r\over (2\pi)^2}{\Lambda^2_k\over \beta_k}
\Big[{\pi\over2}\!-\!2\pi n'\nonumber\\
\!&+&\!i\left({1\over\delta}\!+\!1\!-\!\gamma\!-\!\ln
{\Lambda^2_k\over\mu^2}\right)\Big]-(M=0).
\label{real20}
\end{eqnarray}
In the normal prescription of renormalization of
quantum field theories, we consistently add an appropriate counterterm to cancel the
ultra-violet divergent term $1/\delta$ at the energy-scale $\mu$ in the minimum 
subtraction scheme. 

\section{\it Imaginary part of the effective action.}

We are interested in the imaginary part of the effective action for studying the 
vacuum decay via quantum tunneling through the effective energy gap $m_k$, leading to 
pair-creations of particles and anti-particles. 
Given $\lambda^2_k$ (\ref{ipro0}) and $\Lambda^2_k$ (\ref{lambda0}), we have
\begin{eqnarray}
{\rm Im}\left(\ln {\Lambda^2_k\over\mu^2}\right)\!&=&\! 2\pi n''\!+\!\theta,\hskip0.2cm 
n''=0,\pm 1,\pm 2,\pm 3\cdot\cdot\cdot,
\label{im}\\ 
\theta &= &\tan^{-1}\left[{{\rm Im}
(\Lambda^2_k)\over {\rm Re}(\Lambda^2_k)}\right],
\label{prep}\\
{\rm Im}(\Lambda^2_k)&= &2M\epsilon_k, \label{rla0}\\
{\rm Re}(\Lambda^2_k)&= &\omega^2- \kappa_k^2-m^2_k,\label{ila0}
\end{eqnarray}
where and henceforth the effective mass-gap $m^2_k$ is Eq.(\ref{mb0}) without the angular term 
$l(l+1)/r^2$. Using Eqs.(\ref{epsilon0},\ref{app10}), we define,
\begin{equation}
\Pi_k\equiv {\epsilon_k\over\beta_k}={\int dr g(r)R^*_{l^*\omega}(r)
\hat k_r R_{l^*\omega}(r)\over\int drg(r) |R_{l^*\omega}(r)|^2}.
\label{kapa0}
\end{equation}
As a result, the imaginary 
part of the effective action (\ref{real20}) is given by,
\begin{eqnarray}
{\rm Im}(S_{\rm eff})&=&-\int {d\omega dk_r\over (2\pi)^2}\Big[2M|\Pi_k|
\left({\pi\over2}+\theta-2\pi n\right)\nonumber\\
&+&{{\rm Re}(\Lambda^2_k)\over\beta_k}\left(1-\gamma-\ln{ 
|\Lambda^2_k|\over\mu^2}\right)\Big]-(M=0),
\nonumber\\
n&=& 1,2,3,4,5,\cdot\cdot\cdot,
\label{real30}
\end{eqnarray}
We take the absolute value $|\Pi_k|$ of Eq.(\ref{kapa0}) for both particles and 
antiparticles, as $|k_r|>0$ discussed in the end of section 3. The integer ``$n$'' in 
Eq.(\ref{real30}) describes all 
possible quantum bound states of the vacuum in the presence of negative gravitational potential well.

These quantum bound states ``$n$'' of vacuum in the presence of gravitational potential well
can be understood by an analogue of semi-classical quantization of a particle 
confined in a negative spherical potential $U(r)<0$,
\begin{equation}
\int^{r_c} \! dr|p_r|\! =\!\int^{r_c}\! dr \sqrt{E\!-\!U(r)}\simeq (n\!+\!{1\over2}),
\label{n}
\end{equation}
where $E<0$ is the negative energy of particles.
$p_r$ is the radial momentum conjugated to radius $r$ and integration in Eq.(\ref{n}) 
is limited by the classical point: $U(r_c)=E$.

Quantum-field states tunneling to infinity are free particles and antiparticles 
in mass-shell ${\rm Re}(\Lambda^2_k)=0$. Whereas the quantum-field states $R_{l\omega}(r)$ 
of vacuum in the presence of gravitational field are bound states due to negative
gravitational potential well, this indicates for these states
\begin{equation}
{\rm Re}(\Lambda^2_k)=\omega^2- \kappa_k^2-m^2_k \simeq 0^-,
\label{bn0}
\end{equation}
which we will have more discussions in the last section.
The approximate ``mass-shell condition'' (\ref{bn0}) leads to $\theta\simeq -{\pi\over 2}$ 
in Eq.(\ref{prep}), and gives an approximate dispersion relationship $\omega\simeq\omega(|k_r|)$
between energy $\omega$ and radial momentum $|k_r|$ of states $R_{l\omega}(r)$. 
This allows us to approximately make integration $\int{d\omega dk_r\over (2\pi)^2}$ 
in Eq.(\ref{real30}) with a proper measure $2\pi\delta [{\rm Re}(\Lambda^2_k)]$ in 
the energy-momentum phase space $(\omega,|k_r|)$. As a result, the effective action 
${\rm Im}(S_{\rm eff})$ (\ref{real30}) 
is approximately given by,
\begin{equation}
{\rm Im}(S_{\rm eff})\simeq -
\left(-4\pi M|\Pi_k| n\right).
\label{k0}
\end{equation}
per unit of the number of quantum-field states 
in the range of $\omega\rightarrow\omega +d\omega$.

\section{\it Vacuum decay.}

The imaginary part of $S_{\rm eff}$ determines the probability of vacuum decay caused by 
quantum tunneling of virtual particles:
\begin{equation}
|\langle \tilde 0,{\rm out}|\bar 0,{\rm in}\rangle|^2=e^{-2{\rm Im}S_{\rm eff}},
\label{ie10}
\end{equation}
per unit of the number of quantum-field states in $\omega\rightarrow\omega +d\omega$.   
We sum over all possible quantum states ``$n$'' in Eq.(\ref{real30}) with respect to 
Bose-distribution in the occupation of these quantum states of vacuum. 
The total probability of vacuum decay, leading to pair-creations of real particles and antiparticles, 
is then given as,
\begin{equation}
|\langle \tilde 0,{\rm out}|\bar 0,{\rm in}\rangle|^2 \simeq {1\over
\exp \left(8M\pi |\Pi_k|\right)- 1}.
\label{rate0}
\end{equation}
Regarding these outgoing states of particles as an outward radiation
flux, Eq.(\ref{rate0}) gives the spectrum of such a radiation, with
respect to a static observer located at infinity.  
The characteristic energy scale (temperature) of the spectrum of radiation is 
the Hawking temperature $1/(8\pi M)$. 

As a particular case, we adopt the spherically symmetric solution $l^*=0$ in Eq.(\ref{app0})
and approximate ${2M\over r}\simeq 0$ in the differential equation Eq.(\ref{eq2}) 
so that the spherically symmetric 
solution $R_{0\omega}(r)\sim {e^{ik_rr}\over r}$ is the eigenstate of the radial momentum operators $\hat k_r$ and 
$\hat k^2_r$: 
\begin{eqnarray}
\hat k_r R_{0\omega}(r)& =& k_r R_{0\omega}(r),\nonumber\\
\hat k^2_r R_{0\omega}(r)& = &k^2_r R_{0\omega}(r). 
\label{apph1}
\end{eqnarray}
We approximately have $\kappa_k^2 \simeq k_r^2$ and $m_k^2\simeq m^2$ in
Eqs.(\ref{kr20},\ref{mb0}).
As a result of this approximation and Eq.(\ref{bn0}), we obtain Eq.(\ref{kapa0}):
\begin{equation}
|\Pi_k|\simeq |k_r|\simeq\sqrt{\omega^2-m^2},
\label{apph2}
\end{equation}
and the spectrum of outward radiation Eq.(\ref{rate0}), for $m=0$ massless
radiative field, is the black-body spectrum of the Hawking radiation.

We consider a local observer ${\cal O}$ (\ref{o}) rest at $r^*>2M$. With respect to this local observer, we introduce 
a spherical shell whose radial size $\Delta r\ll r^*$, indicating 
the variation of the external gravitational field is very small within the shell. 
We can approximately treat the external gravitational field as a constant field in 
the shell. On the other hand, $\Delta r$ is much larger than the infrared
cutoff of quantum field theories. Given approximate constancy of gravitational field 
in the shell, we can approximately make substitution $r\rightarrow r^*$ and $\int dr r^2\rightarrow \Delta r (r^*)^2$,
in the definitions $N_k$ (\ref{norma0}), $m^2_k$ (\ref{mb0}) and $\beta_k$ (\ref{app10}). 
$k_r$ (\ref{kr0}), $\kappa^2_k$ (\ref{kr20}), and $\epsilon_k$ (\ref{epsilon0}) become: 
\begin{eqnarray}
k_r(r^*) &=& \left[{R^*_{l\omega}(r)\hat k_r R_{l\omega}(r)\over |R_{l\omega}(r)|^2}\right]_{r=r^*},\label{kr}\\
\kappa^2_k (r^*)&=& g^2(r^*)\left[{R^*_{l\omega}(r)\hat k^2_r R_{l\omega}(r)\over |R_{l\omega}(r)|^2}\right]_{r=r^*};
\label{kr2}\\
\epsilon_k (r^*)&=& {g(r^*)\over (r^*)^2}\left[{R^*_{l\omega}(r)\hat k_r R_{l\omega}(r)\over 
|R_{l\omega}(r)|^2}\right]_{r=r^*}.
\label{epsilon}
\end{eqnarray}
With respect to this local observer, the effective mass-gap $m_k$ (\ref{mb0}) is, 
\begin{equation}
m^2_k(r^*)= g(r^*)V_l(r^*),\label{mb}
\end{equation}
and the approximate ``mass-shell condition'' (\ref{bn0}) becomes
\begin{equation}
{\rm Re}(\Lambda^2_k)(r^*)=\omega^2- \kappa_k^2(r^*)-m^2_k(r^*) \simeq 0^-,
\label{bn}
\end{equation}
where $\kappa_k^2(r^*)$ is given by Eq.(\ref{kr2}) and $m^2_k(r^*)$ by Eq.(\ref{mb}).
And $\Pi_k$ (\ref{kapa0}) becomes,
\begin{equation}
\Pi_k(r^*) = \left[{R^*_{l^*\omega}(r)
\hat k_r R_{l^*\omega}(r)\over |R_{l^*\omega}(r)|^2}\right]_{r=r^*}=k_r(r^*).
\label{kapa}
\end{equation}
Analogously to Eq.(\ref{rate0}), the probability of vacuum decay via pair-production is given by,
\begin{equation}
|\langle \tilde 0,{\rm out}|\bar 0,{\rm in}\rangle|^2 \simeq {1\over
\exp \left(8M\pi |\Pi_k(r^*)|\right)- 1},
\label{rate}
\end{equation}
where $\Pi_k(r^*)=k_r(r^*)$ (\ref{kapa}) and $k_r(r^*)$ is determined by Eq.(\ref{bn}).
For spherically symmetric case $l^*=0$ (\ref{apph1}), $r^*\gg 2M$, $m=0$ and $|k_r(r^*)|\simeq \omega$, 
the probability of vacuum decay with respect to the local observer is 
\begin{equation}
|\langle \tilde 0,{\rm out}|\bar 0,{\rm in}\rangle|^2 \simeq {1\over
\exp \left(8M\pi |k_r|\right)- 1} \simeq {1\over
\exp \left({\omega_{\rm loc}\over T_{\rm loc}}\right)- 1},
\label{ratelocal}
\end{equation}
where $\omega_{\rm loc}$ and $T_{\rm loc}$ are the energy-spectrum and temperature: 
\begin{equation}
\omega_{\rm loc}=g^{-{1\over2}}(r^*)\omega,\hskip0.5cm 
T_{\rm loc}={1\over8\pi M}g^{-{1\over2}}(r^*),
\label{local}
\end{equation}
with respect to the local observer rest at $r^*$. 
$\omega_{\rm loc}$ and $T_{\rm loc}$ with 
respect to the local observer ${\cal O}$ are gravitationally red-shifted from their counterparts
$\omega$ and $T$ with respect to the static observer at infinity.

It is worthwhile to compare the some results of this article with the WKB-method adopted 
in the brick wall model\cite{thooft,brick}, where the solution to Eq.(\ref{eq2}) in 
the vicinity of black horizon was found. The radial wave-vector $k_r$ (\ref{kr0}) or 
$k_r(r^*)$ (\ref{kr}) we define is consistent with that given in refs.\cite{thooft}. 
Actually, the value of the radial wave-vector $k_r(r^*)$ determined by Eq.(\ref{bn}) is the exactly same as the wave-vector (8.9) obtained by directly solving Eq.(\ref{eq2}) 
in refs.\cite{thooft}. 
The oscillatory behaviour of the quantum field $\phi$ (8.7) in refs.\cite{thooft} 
for describing quantum tunneling by the WKB-method coincides with the imaginary part of effective action (\ref{k0}) obtained in this article. The total number $\nu$ of radial 
wave-solution up to a given energy, discussed in refs.\cite{thooft}, is equivalent 
to the radial quantum number ``$n$'' (energy-state) discussed in Eqs.(\ref{real30},\ref{k0}). 
In both schemes, these states ``$n$'' are occupied by non-negative number of quanta $|k_r|$ to give rise to the energy-spectrum of quantum radiation.

\section{\it Discussions.}

We turn to discussions of the dynamical reason for such a vacuum decay and
particle creations. In the framework of quantum field theories, quantum-filed fluctuations in vacuum
indicate pair creations and annihilations of positive- and negative-energy 
virtual particles, represented by closed fermion-loops in Feynman
diagrams. We point out that in the presence of external gravitational field, such an pair-creation process is 
energetically favorable for it gains a gravitational energy $\delta E<0$. Setting $\delta r\ll r^*$ is the quantum 
separation of positive- and negative-energy virtual particles in the radial direction, $\delta t$ and $\delta\omega$ 
are time and energy variation of the pair-creation process, we can approximately compute the gravitational energy gain $\delta E$ per quantum state $2\pi$:
\begin{eqnarray} 
\delta E\!&= &\!g^{1\over2}(r^*\!-\!{|\delta r|\over2}){|\delta\omega_{\rm loc}|\over2\pi}
\!-\!g^{1\over2}(r^*\!+\!{|\delta r|\over2}){|\delta\omega_{\rm loc}|\over2\pi}\label{pn}\\
&\simeq &\!-\!{M\over 2\pi (r^*)^2}|\delta\omega||\delta r|g^{\!-\!{1\over2}}(r^*)
\simeq\!-\! {M\over 2\pi (r^*)^2}g^{\!-\!{1\over2}}(r^*),
\label{energygain}
\end{eqnarray}
where the first and second terms in Eq.(\ref{pn}) are gravitational energies respectively for positive ($\delta \omega_{\rm loc} >0$) and 
negative energy ($\delta \omega_{\rm loc} < 0$) virtual particles. 
$\delta\omega_{\rm loc}=g^{-{1\over2}}(r^*)\delta\omega$ is 
the energy variation in a local rest frame of the static observer ${\cal O}$ (\ref{o}) at $r^*$. 
In Eq.(\ref{energygain}), we adopt 
$|\delta\omega|\simeq |\delta k_r|$  and the Heisenberg uncertainty relationships:
\begin{equation}                                   
\delta t\delta\omega\simeq 1;\hskip0.5cm \delta r\delta k_r\simeq 1,
\label{hu}
\end{equation}
which are invariant in any arbitrary coordinate frames. Analogously, given the gravitational potential $-{M\over r^*}$ 
in the Newtonian limit and $\delta\omega \simeq\delta\omega_{\rm loc}$, we have
\begin{eqnarray} 
\delta E\!&\simeq &\!-{M{|\delta\omega_{\rm loc}|\over2\pi}\over 
r^*\!-\!{|\delta r|\over2}}\!+\!{M|{|\delta\omega_{\rm loc}|\over2\pi}\over r^*\!+\!{|\delta r|\over2}}\nonumber\\
\!&\simeq &\!-{M\over 2\pi (r^*)^2}|\delta\omega||\delta r|\!\simeq\!-{M\over 2\pi (r^*)^2}.
\label{nenergygain}
\end{eqnarray}
This shows that the gravitational field polarizes the vacuum by displacing 
$-{|\delta r|\over2}$ for positive energy virtual particles and ${|\delta r|\over2}$ 
for negative energy virtual particles, such a displacement gains gravitational energy. 
As a consequence of gravitational vacuum polarization, vacuum possibly 
decays via pair-creation process to create real particles and anti-particles, leading to 
quantum radiation in curved spacetime. 
This is analogous to the phenomenon of an external electric field polarizing vacuum,
leading to possible pair-creations of particles and anti-particles, as described by the 
Schwinger mechanism\cite{sw}.

This gravitational vacuum polarization is characterized by the gravitational 
energy-gain $|\delta E|$ (\ref{energygain}) with respect to the local observer at $r^*$. 
In the proper frame of a free-falling observer, whose acceleration
$a=g^{\!-\!{1\over2}}(r^*) {M\over (r^*)^2}$ at $r^*$, the energy-gain is, 
\begin{equation} 
|\delta E |=\!{M\over 2\pi (r^*)^2}g^{\!-\!{1\over2}}(r^*)={a\over2\pi}.
\label{uenergygain}
\end{equation}
This is reminiscent of the Unruh effect\cite{ueffect} and indicates that its origin 
has the same dynamical nature of gravitational vacuum polarization. An accelerating 
observer finds gravitational vacuum polarization leading to pair-creations in his/her proper frame, 
as required by the equivalent principle.
 
With respect to an infinity observer, the gravitational energy-gain for pair-creations 
is ${M\over 2\pi (r^*)^2}$, which can be found in Eq.(\ref{energygain}) consistently with 
Eq.(\ref{local}). In the neighborhood of an eternal black hole horizon
($r^*\rightarrow 2M$), the gravitational energy-gain is maximum and in fact 
determines the Hawking temperature $1/(8\pi M)$, which is very small, compared with 
neutrino masses $m_\nu$. However,  
the effective mass-gap $m_k(r^*)$ (\ref{mb}) vanishes in the vicinity of black hole's horizon. 
As the result of this vanishing effective mass-gap,
virtual particles that are in quasi zero-energy states $\omega\sim
0^-$ just bellow the zero-energy level of vacuum, by quantum-field fluctuations, 
turn to be real particles that are in quasi zero-energy states $\omega\sim 0^+$ 
just above the zero-energy level of vacuum, 
since it almost costs no energy for such a quantum
tunneling process crossing the zero-energy level of
vacuum and leading to pair-creations. These particles in quasi zero-energy 
states have a typical energy-scale ${M\over 2\pi (r^*)^2}$ ($r^*\rightarrow 2M$), 
$~{1\over 8\pi M}$ the Hawking temperature. We emphasize three properties
of particle pair-creation in the vicinity of black hole horizon: (i) gravitational 
energy-gain is maximum; (ii) the effective mass-gap of virtual particles in the vacuum
is vanishing; (iii) these lead to the largest probability of particle pair-creations,
the rate and spectrum of the Hawking radiation (\ref{ratelocal}).

Due to the smallness of the effective mass-gap $m_k\sim 0$, real particles and 
antiparticles created, are in the mass-shell condition
$\omega =|k_r|$, which are quasi zero-energy states $\omega\sim 0^+$. By the 
continuation of energy-momentum
dispersion relation, virtual particles and antiparticles, which are in quasi 
zero-energy states $\omega\sim 0^-$, must be
in an approximate mass-shell condition $\omega\simeq -|k_r|\sim 0^-$. This justifies 
the ``mass-shell condition'' ${\rm Re}(\Lambda^2_k)\simeq 0^-$ (\ref{bn0}) and (\ref{bn}).
${\rm Re}(\Lambda^2_k)<0$ is due to virtual particles bound by
negative gravitational potential. 

In the region away from black hole's horizon ($r^*>2M$) and massless particles $m=0$,
the effective mass-gap $m_k(r*)\sim \left({2M\over (r*)^3}\right)^{1\over2}$ is extremely
small and comparable with the gravitational energy-gain ${M\over 2\pi (r^*)^2}$. 
Pair-creation rate is given by Eq.(\ref{rate}) in very low-energy regime 
($\sim 1/r^*$). This could be 
case for extremely low-energy emission of massless particles. With respect 
to an infinity observer, the characteristic energy of such low-energy emissions is given by
\begin{equation} 
T=|\delta E |={M\over 2\pi (r^*)^2}={1\over8\pi M}\left({2M\over r^*}\right)^2,
\label{awayt}
\end{equation}
which is smaller than the Hawking temperature ${1\over8\pi M}$. The power of 
such low-energy emissions (black body radiation) is 
\begin{equation}
P=4\pi (r^*)^2 \sigma_B T^4,
\label{powerh}
\end{equation}
where $\sigma_B$ is the Stefan-Boltzmann constant. The life-time of a gravitational body $M$, 
describing its instability against this radiation,
is approximately given by
\begin{equation}
\tau\simeq {M\over P}.
\label{lifeh}
\end{equation}
We examine two examples: the Earth ${2M_\oplus\over r_\oplus}=7\cdot 10^{-10}$ where 
$M_\oplus$ is Earth's mass and $r_\oplus$ Earth's radius, a proton 
${M_p\over r_p}=8\cdot 10^{-37}$ where $M_p$ is proton's mass and $r_p$ proton's 
classical radius. We find the temperature, power and life-time of the Earth ($\oplus$)
and a proton ($p$):  
\begin{eqnarray}
T_\oplus &\simeq& 4\cdot 10^{-20}K^\circ,\hskip 0.3cm P_\oplus\simeq 7\cdot 10^{-65}
{\rm ergs/sec},\hskip 0.3cm \tau_\oplus\simeq 2\cdot 10^{85}{\rm years};\label{earth}\\
T_p &\simeq& 2\cdot 10^{-22}K^\circ,\hskip 0.3cm P_p\simeq 3\cdot 10^{-122}
{\rm ergs/sec},\hskip 0.3cm \tau_p\simeq 2\cdot 10^{111}{\rm years}.
\label{proton}
\end{eqnarray}
Compared with the life-time of the Universe $\sim 10^{10}$ years, these show that 
gravitational bodies are stable against such quantum radiation attributed to 
gravitational vacuum-polarization. On the other hand, all known 
fermion masses ($m$) are not zero and much larger than $\sim 1/r^*$. As a consequence, 
the rate (\ref{rate}) of quantum tunneling effect (quantum radiation) 
is exponentially suppressed ($\sim e^{-8\pi M m}\simeq 0$).

Attributed to the nature of quantum-field fluctuations of virtual
particles tunneling through a very small effective mass-gap, 
vacuum decays, leading
to pair-creations of real particles and antiparticles. Accordingly,
the nature of quantum-field fluctuations of real particles and
anti-particles, clearly
implies the inverse process: pairs of real particles and antiparticles
annihilate into virtual particles and antiparticles in vacuum. 
The rate of pair-annihilation process must be the same as the rate of 
pair-creation process, as the CPT invariance is preserved in such
processes. In these emission(creation) and absorption(annihilation) 
processes, the vacuum acts as a reserve of temperature given by the 
gravitational energy-gain Eq.(\ref{energygain}). In the presence of gravitational 
field, the spectrum of this reserve(vacuum) is quantized and described by the integer $n$ 
and quanta $|k_r|$ (see Eqs.(\ref{real30},\ref{kapa})). Particle creations(emissions) from vacuum and 
annihilation(absorptions) into vacuum are permitted, 
if these processes take place in unit of quanta $|k_r|$. The detail balance of
these emission and absorption processes in unit of quanta $|k_r|$,
leads to the black-body spectrum (\ref{rate}) of particle creations.
In general, such a spectrum is different from black-body one up to a gray factor, since 
${\rm Re}(\Lambda^2_k)$ is not exactly zero for virtual particles 
in quasi zero-energy states ($\omega\sim 0^-$) and the approximation (\ref{app0}) is made. 
This is actually because of particle creation and annihilation 
processes scattered by the potential terms ${2M\over r^3}$, $l(l+1)/r^2$ and $g(r)$ 
in the effective mass-gap. 

In this article, we discuss that with respect to a static local observer 
at $r^*<\infty$ and a static observer at infinity, the transition amplitude 
between the vacuum state $|\tilde 0,{\rm out}\rangle$ (\ref{vd}) of curved spacetime 
and the vacuum state $|\bar 0,{\rm in}\rangle$ (\ref{vd1}) of flat spacetime, showing 
the vacuum state $|\tilde 0,{\rm out}\rangle$ is unstable against
the vacuum state $|\bar 0,{\rm in}\rangle$. This is attributed to the nature 
of gravitational field polarizing the vacuum state $|\bar 0,{\rm in}\rangle$ and 
such polarizations gaining gravitational energy, leading to particle and antiparticle
productions. As a consequence, the static observer ${\cal O}$ (\ref{o}) 
in curved spacetime measures the vacuum state $|\tilde 0,{\rm out}\rangle$ (\ref{vd}) 
as thermal radiation of particle and antiparticle productions upon the vacuum state
$|\bar 0,{\rm in}\rangle$ (\ref{vd1}). Although such an effect 
is in general extremely small (\ref{earth},\ref{proton}), it converts the gravitational 
energy (matter) into the energy of radiation fields.  

\newpage



\end{document}